\documentclass[twocolumn,english]{autart}

\pdfoutput=1
\usepackage[utf8]{inputenc} 
\usepackage{moreverb}
\usepackage{graphicx}
\usepackage{amsmath}
\usepackage[wide]{sidecap}
\usepackage{lipsum,babel}
\usepackage{setspace}

\begin{document}

\begin{frontmatter}
	
	\title{Power Quality Monitor for Residential Voltage}
	\author{Daniel Pérez}\ead{daniel.andres.perez.espina@uru.edu}
	\address{School of Electrical Engineering, Engineering Faculty, Universidad Rafael Urdaneta, Maracaibo, Zulia, Venezuela}
	
	\begin{abstract}
	{A simple and inexpensive system capable of displaying relevant power quality data of a residential voltage signal is presented, and it is based on Arduino, Python and Kivy. A circuit for adjusting the voltage signal for sampling is suggested. The software of the system allows the transmission of data with any device capable of serial communication at a 2000000 Baud rate, although an Arduino NANO/UNO board is recommended. A board based on an Arduino NANO R3 was used. The software plots the last 6 cycles contained in the data by default, but lets analyze any amount of cycles. It provides the RMS voltage, the peak voltage and the total harmonic distortion of the scrutinized cycles, up to the 25th harmonic. The software is also capable of calculating the fast Fourier transform of the studied cycles, and returning the corresponding plot. A sample of data captured at Maracaibo, Venezuela is shown.}
	\end{abstract}
\end{frontmatter}

\section{Introduction}
The current state of Venezuela's electrical system is extremely dire, due to a long and ongoing period of mismanagement \cite{WSJ}. All the monitoring, maintenance and expansion that is required to sustain the system hasn't been done in a considerable amount of years. The consequences of the mismanagement can be summed up as the decay of the system, which has brought many power disturbances such as voltage sags and surges, being these two disturbances a prominent source for damage of electrical appliances.\\
For the past few years, no data on the topic has been pu-blicly disclosed by official sources. Although it is agreed among many Venezuelans that the electrical service is in bad shape, it is extremely hard to make an objective analysis of its current state without using any sort of data. Taking into account that there's no available information on the subject, the only possible way to get data about Venezuela's electrical system is to use power qua-lity monitoring equipment, but being that Venezuela's economy is in an appalling state, this option turns out to be extremely expensive for locals. Gathering data city-wide is desirable, but making a full scale analysis of all the distribution circuits of an entire city would require a significant amount of equipment.\\
Thanks to the rise of micro-controllers in the modern era, it is possible to access low-cost devices capable of doing analog to digital conversions, which are required for gathering the data of the electricity. A simple and inexpensive system that works as a solution to this problem is showcased in this paper. Said system is able to continuously sample, measure the RMS value of the voltage, the peak voltage, and the total harmonic distortion (THD) considering up to the 25th harmonic.\\
The system is comprised by two parts: the data gathe-ring part, which uses its own piece of software and an Arduino board alongside a circuit; and the data analysis part, which processes and displays in a computer all the gathered data. To achieve a low cost approach, an Arduino-based board is used alongside a few spare parts for a circuit, and a Dell Vostro 3500 with Linux, Python and Kivy. The reasoning behind these selections is that they represent free software paired with very low cost hardware. This combination is able to capture, process and analyze data from a residential voltage signal, which in Venezuela's case is 120 VRMS at 60 Hz \cite{CEN}. The overall goal of this system is to make a scalable solution that could be used for making an analysis of the status of all the distribution circuits of Maracaibo, Venezuela.

\section{Data Gathering}
For the design of the system a board based on the Arduino NANO R3 was used, due to its low cost. Under this setup an Arduino NANO or UNO can be used, or just an ATmega 328P for that matter. Other MCU devices can be used, but that will deviate the system from what is presented in this paper.\\
It was established that the workload of the Arduino board should be kept at a minimum, and that a computer must be used to do the signal processing. This decision was not only based on the low processing power of the ATmega 328P, but in the fact that increased CPU activity of the micro-controller leads to generation of noise \cite{AVR}, which is undesirable. There were a lot measures taken for ensuring the lowest amount of noise possible (excluding shielding the device), and they comprise the following:
\begin{itemize}
  \item Setting all unused pins as inputs.
  \item Connecting all unused pins to ground.
  \item Utilizing the noise reduction mode for doing analog readings.
  \item Disabling the digital input buffer of analog pins.
  \item Using decoupling capacitors for the analog reference and power pins.
  \item Not using target computer as source of power.
  \item Keeping analog and power lines as close as possible.
\end{itemize}
These measures were based on the fact that floating pins generate noise \cite{CDC}, and on various EMC, grounding and power electronics concepts \cite{GND},\cite{EMC},\cite{EMG},\cite{VR}. Any unused components of the ATmega 328P were disabled. Although the following wasn't achieved for this system, isolating the Arduino board as much as possible from the target computer will reduce the electrical noise sent to the board by the computer. The best method of isolation is avoiding any electrical connections whatsoever.\\
To stay in compliance with the Nyquist sampling theorem \cite{DSP}, a 3600 Hz sampling rate was chosen. This sampling rate not only lets capture the data of the fundamental frequency of the residential voltage signal, which is 60 Hz, but it also lets capture data for all harmonics of said signal up to the 25th one.\\
For the presented scenario the Nyquist frequency is 1500 Hz, which means that a 3000 Hz or higher sampling rate is required. Although increasing the sampling rate is possible and would in theory provide higher quality data, that would also require increasing the clock frequency of the analog to digital converter (ADC), which would lead to reducing the resolution of the ADC. The default frequency of the ADC clock for the Arduino board is 125 kHz. Any value above 200 kHz makes 10-bit readings unreliable, forcing to use 8 bits-only readings \cite{AVR}. Due to the way the prescalers of the ADC and the system clock frequency work, it is not possible to set the ADC clock frequency to any value between 125 kHz and 200 kHz.\\
Between the limitations of the ADC of the ATmega 328P, lies the input restriction of 0 - 5 V. This implies that signal manipulation is required in order to sample the residential voltage. Considering that potential transformers are a source of noise and could ramp up the overall cost of the system, the use of a voltage divider with resistors was preferred and is recommended. This is acceptable due to the internal resistance of the ADC being 100M$\Omega$\cite{AVR}, being that a set of values that are significantly lower can be used. Alongside the voltage divider, an external DC source of 3.3 V was used to provide an offset to the signal, leaving the following equivalent circuit:\\
\begin{figure}[!ht]
\centering
\includegraphics[width=85mm]{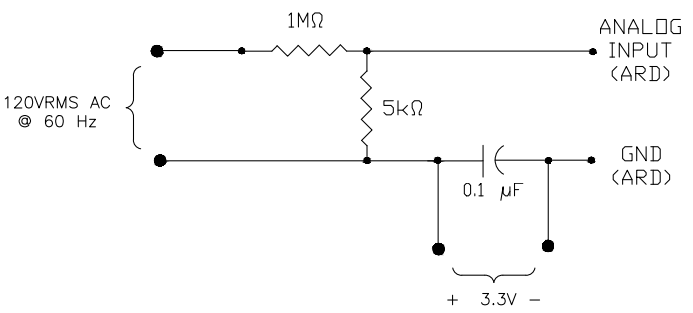}
\caption{Circuit that outputs signal with 3.3 V offset and .5\% amplitude from input}
\label{circuit}
\end{figure}
\\The main takeaway from figure \ref{circuit} lies in the fact that the power losses related to the voltage divider are very low ($\simeq$ 14.33 mW), so there's no true significant advantage from using a transformer for this particular part. Based on the circuit in figure \ref{circuit} it is possible to create a simple expression for the ratio of the voltage in the 5 k$\Omega$ resistor in relation to the residential voltage:
\begin{equation}
\frac{V_{o}}{V_{i}} = \frac{5 k\Omega}{1 M\Omega + 5 k\Omega} \simeq 0.0050
\label{eq:out}
\end{equation}
Under this configuration the 5 k$\Omega$ resistor perceives approximately 0.5\% of the residential voltage at any time, but it is to be taken into account that this value may vary if the resistors experiment temperature changes \cite{CDC}. This implies that for a 120 VRMS signal there's an output of 0.6 VRMS with the same properties as the input. Said output has a set of peak values of $\pm 0.6\times\sqrt{2}$ V, which transgresses the input limit of the ADC, being that there's no tolerance for negative values. Due to this fact it was decided to use a 3.3 V source for offsetting the output signal from the voltage divider, making the peak values of the resulting signal 3.3 V $\pm 0.6\times\sqrt{2}$ V, which are both positive values under 5 V. This confi-guration also provides great tolerance against voltage surges, being that an increase of up to 100\% in amplitude would still produce an upper peak under 5 V, and a voltage surge of 100\% or higher is never to be expected in Maracaibo \cite{FIDVR}. The voltage provided by the 3.3 V DC source was filtered with a decoupling capacitor.\\
It is worth noting that although modern regulations establish that for the THD calculation all the harmonics up to the 50th should be considered \cite{519}, the aim of this device isn't being equal to professional power quality equipment, the main goal is to give insight about the current state of Venezuela's electrical service based on Maracaibo's data, as there is no available information at a research level. It is possible to set the ATmega 328P to sample at the 6000 Hz required for registering up to the 50th harmonic without modifying the prescalers, but that would require scrapping the use of the Arduino C/C++ functions, and take an approach directly focused in plain C or Assembly. This approach is planned to be added in the future.\\ 
The code required for sampling the resulting signal and then sending it to the target computer is very short and simple. A snippet of the code that demonstrates the whole process from a high level perspective, can be seen here:\\
\begin{listing}{20}
void setup() {
  setAllPinsAsInput(); // Reduces noise
  analogRead(A2); // Set ADMUX, initial read
  deactivateDigitalInputs(); // Reduces noise
  Serial.begin(2000000); }

void loop() {
  StartTime = micros();
  value = NRADC(); // ADC with noise reduction
  sprintf(buffer,"
  Serial.println(buffer);
  while (micros() < StartTime + 162); }
\end{listing}
The entire code only has 48 lines, and is represented overall by the prior snippet. The use of a buffer as opposed to a String data type is due to preventing me-mory allocation issues, which is a prominent problem with C/C++ coding in general \cite{MESH}. Using the String data type in the dynamic memory of a micro-controller can lead to program crashing under continuous use, due to memory fragmentation.\\
The data is sent with a Baud rate of 2000000, which isn't strictly necessary but it ensures that each reading takes under approximately 277 $\mu$s to be sent, and therefore the sampling rate is very close to 3600 Hz. The use of an Arduino board for gathering and sending the data isn't required, but it opens up the option of making a low cost Internet of Things infrastructure \cite{ARD}, which could be used for sending to the cloud the data of all the distribution circuits of an entire city, such as Maracaibo, in order to make a deep analysis.
\section{Data Processing}
For this section, it was decided to work with Python and Kivy. The Matplotlib, NumPy and SciPy modules are able to handle all the operations that the system is required to make. The NumPy module is only used as an auxiliary tool for plotting with Matplotlib and making operations related to the THD. The SciPy module is exclusively for getting the fast Fourier transform (FFT), which is used both for calculating the THD and plotting the FFT of any amount of cycles of the sampled voltage signal. The PySerial module is used to establish the serial connection and receive each line of data. It is compatible with the 2000000 Baud rate \cite{Ser}.\\
All the raw data gets stored in a file called ``dataRaw.bin'', and is represented by integers separated by a comma (``,''). At the beginning of each transmission of data, the exact date where the process started is printed to the file, as a way to keep track of the time of transmission. The number of values stored can be easily obtained, and based on the time taken to send each value, it can be determined the total transmission time. The stored values are integers returned by the ADC of the ATmega 328P in the Arduino board, and range between 0 and 1023, where 1023 represents the value of the analog \mbox{reference} \cite{AVR}. It is paramount to establish exactly what the analog reference voltage is, since it is required to convert the stored values into instantaneous voltages, which are needed for all the operations that the software has to make. Unless an external analog reference is used, said value is always going to be the same as the one being used to power the system. This is the expression utilized for converting the values sent by the Arduino board:
\begin{equation}
  Voltage = \frac{\left(Value\times\frac{Vcc}{1023}\right) - 3.3 V}{0.005}
  \label{eq:transform}
\end{equation}
As may be hinted in \ref{eq:transform}, the values of $Vcc$ and the voltage source used for offsetting can be a source of error. Cons-tant monitoring of their effects due to variation have to be taken into account when making an interpretation of the data. The function returns the equivalent amount of values to an integer number of cycles, not to leave any data behind and to avoid spectral leakage \cite{DSP}. An option to retrieve only complete cycles will be added in the future.\\
The code used for calculating the RMS value and the peak voltage is simple. The RMS voltage returned is based on a discrete calculation using the given sample, while the peak voltage returned is the highest absolute value between all the given data. The code in question is the following:\\
\begin{verbatim}
def getVRMS(values):
  total = 0
  for each_item in values:
    v2 = each_item**2
    total += v2
  VRMS = (total/(len(values)))**(1/2)
  return VRMS

def getVP(values):
  return max(max(values),abs(min(values)))
\end{verbatim}
The expression utilized for the values of the FFT, which is used for plotting and getting the THD, is the following:
\begin{equation}
  \hat{V} [k] = \sum_{n=0}^{n\times 60} e^{-2\pi j\frac{kn}{n\times 60}} V[n]
\label{eq:fft}
\end{equation}
For this expression both $V$ and $\hat{V}$ have $n\times 60$ values, which represents the amount of values within $n$ cycles with a sampling rate of 3600 Hz. The FFT algorithm of SciPy returns all the positive values in the first half \cite{FFT} of $\hat{V}$, while it returns all negative values in the remaining half. For both plotting the FFT and calculating of the THD, only the modulus of the values of the transform are used. The FFT returned by SciPy under this scenario covers up to 1800 Hz, which is over the 1500 Hz of the 25th harmonic.
The expression used for calculating the THD is the following:
\begin{equation}
  THD = \frac{\sqrt{\sum_{n=2}^{25}{\left|\hat{V}_{n}\right|}^{2}}}{{\left|\hat{V}_1\right|}}
\end{equation}
Where $\hat{V}_n$ represents the voltage of each $n$ harmonic in the frequency domain, and $\hat{V}_1$ represents the voltage of the fundamental harmonic of the signal.\\
It is important to consinder that the FFT algorithm of SciPy always truncates the input signal \cite{FFT}. Said truncation can be modeled as the multiplication of a corres-ponding infinite signal with a window function. This translates as a convolution in the frequency domain \cite{DSP}, which leads into spectral leakage. This effect inherently introduces a deviation in the THD calculation, being that there's going to be values above zero for all frequencies in $\hat{V}$. Nonetheless, the behavior of the spectral leakage for an ideal discrete sine wave can be used to identify whether or not there's a presence of harmonics in the sampled signal. The FFT of a discrete sine wave of 60 Hz, 6 cycles and 60 values per cycle can be seen in the following figure:
\begin{figure}[!ht]
\centering
\includegraphics[width=90mm]{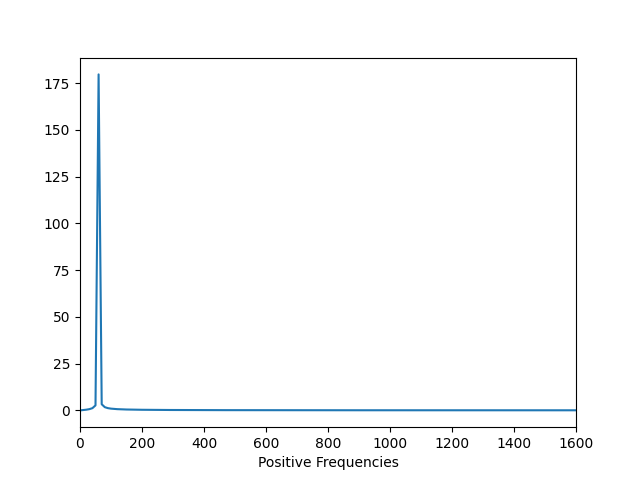}
\caption{FFT of discrete ideal sine wave}
\label{idealsine}
\end{figure}
\\As can be observed in figure \ref{idealsine}, after the expected spike at 60 Hz the values started to drop. This means that for there to be energy transmission at any other positive frequency, there has to be another spike that will break and reintroduce the trend of decaying values. By adding a similar discrete sine wave of amplitude $\frac{1}{2}$ and 180 Hz to the original sine wave, we get the following behavior:
\begin{figure}[!ht]
\centering
\includegraphics[width=90mm]{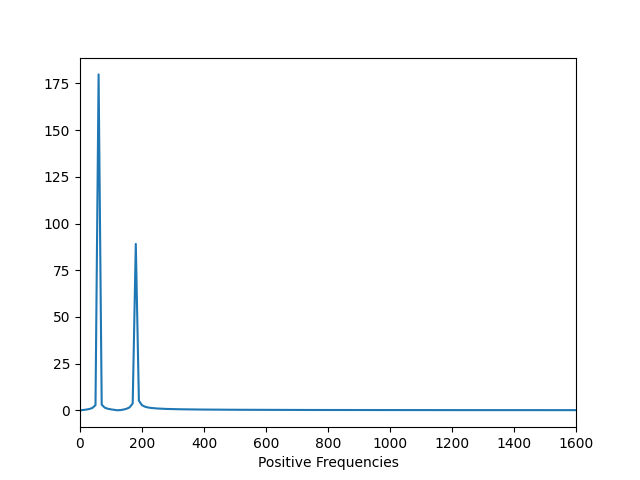}
\caption{FFT of ideal discrete sine wave and 3rd harmonic}
\label{sinenh}
\end{figure}
\\As previously asserted, the introduction of a harmonic broke the lowering pattern due to a spike, and after said spike the pattern returned. Due to this fact alone, it was decided only to take into account the voltages of harmonic frequencies that were higher than the prior and following value, in such way that only spikes can be considered for the calculation. This negates the values on any harmonic frequencies that don't meet this criteria, and therefore their contribution to the calculation of the THD is none.\\
In the following figure can be seen the GUI for establi-shing connection with the target PC, and start gathe-ring data:
\begin{figure}[!ht]
\centering
\includegraphics[width=65mm]{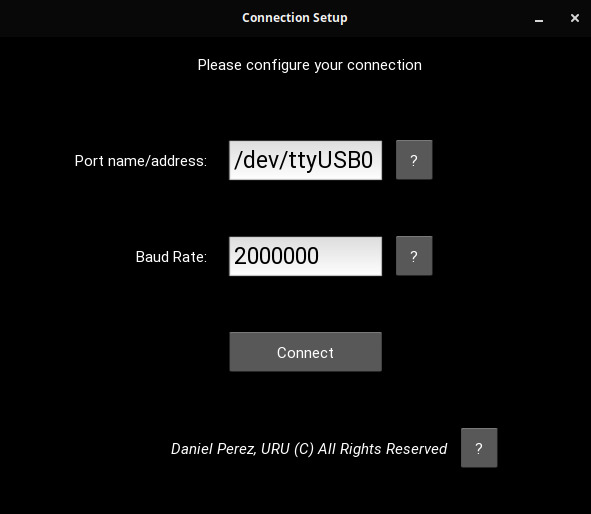}
\caption{Connection GUI}
\label{CGUI}
\end{figure}
\\This GUI is its own piece of software, and it is designed to be simple, intuitive and consume the least amount of resources possible. Once the connection is established, the Kivy environment that runs the program freezes and only the data storing script is left working in the background.
\newpage In figure \ref{CGUIConnected} can be seen the connection screen:
\begin{figure}[!ht]
\centering
\includegraphics[width=65mm]{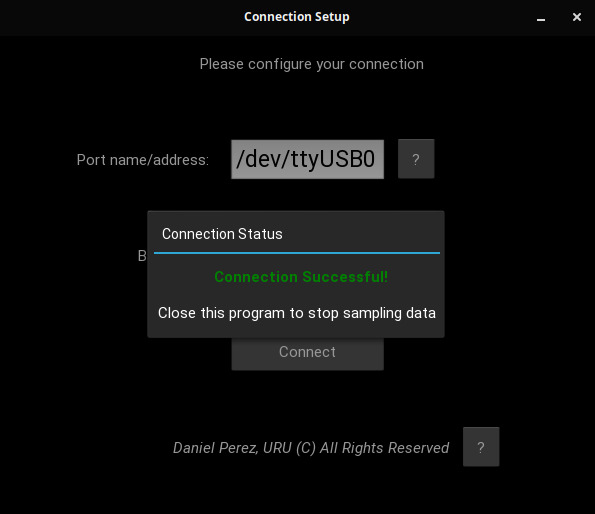}
\caption{Connection status in GUI}
\label{CGUIConnected}
\end{figure}
\\This system has a second GUI, which serves as the main one and it is where all the data gets displayed. The main GUI has its own code that processes the stored data. The main GUI can be seen in action in figure \ref{mainGUI}:
\begin{figure}[!ht]
\centering
\includegraphics[width=85mm]{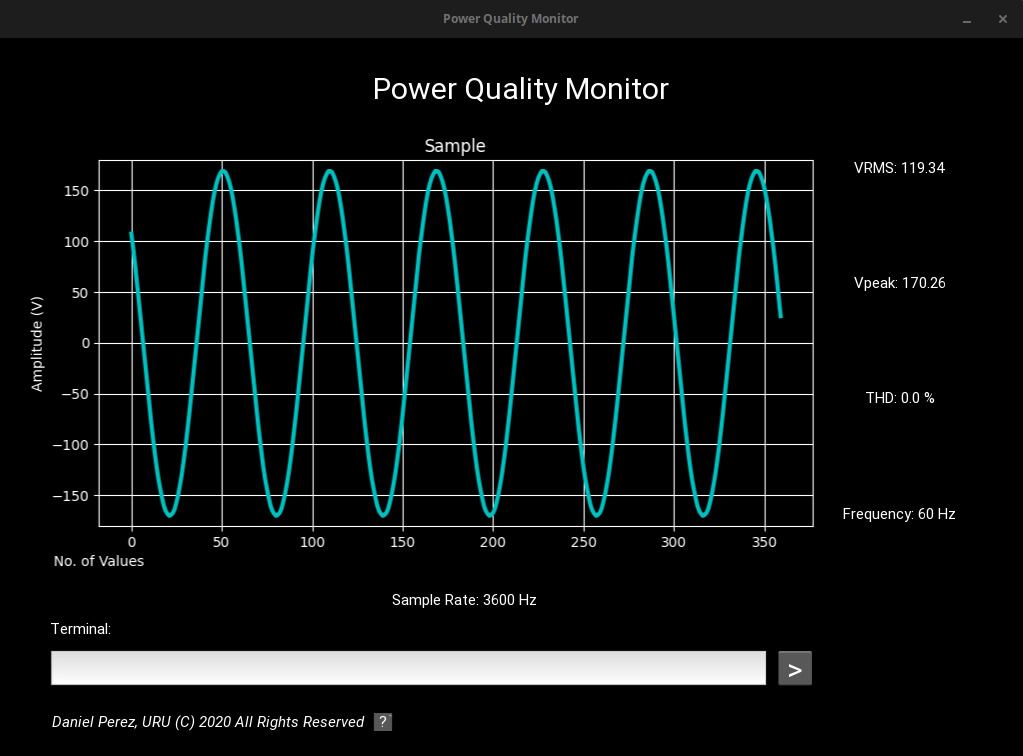}
\caption{Main GUI (with mock data)}
\label{mainGUI}
\end{figure}
\\The sample gets plotted using Matplotlib, and the data displayed on the right side is calculated based on the values used for plotting. The frequency displayed is only referential and not actually calculated, although this is planned to be added in the future. The main GUI can work as long as there's stored data, regardless there's transmission happening or not. It is capable to analyze and make FFT-related operations to any range of cycles.\\
The main GUI also has the ability of converting the data of the ``dataRaw.bin'' file into RMS $\frac{1}{2}$ values, after which it stores them in a file named ``dataRMS.bin''. This file is extremely lighter due to the nature of its data, an RMS $\frac{1}{2}$ value takes the place of 30 instantaneous va-lues. The RMS $\frac{1}{2}$ values can be displayed just like the instantaneous values. The code of the main GUI can also perform a power quality analysis to the values in ``dataRMS.bin'', and give a full report. The main stats of the report are shown in a popup inside the main GUI, but a .txt file containing an in-depth analysis of all perturbations is produced as well, and is stored locally as ``Report.txt''
\section{Analysis of Sample}
The last 6 cycles of a sample taken on Maracaibo, Venezuela on May 6th of 2020, at 4:01 P.M. are in the following figure:
\begin{figure}[!ht]
\centering
\includegraphics[width=85mm]{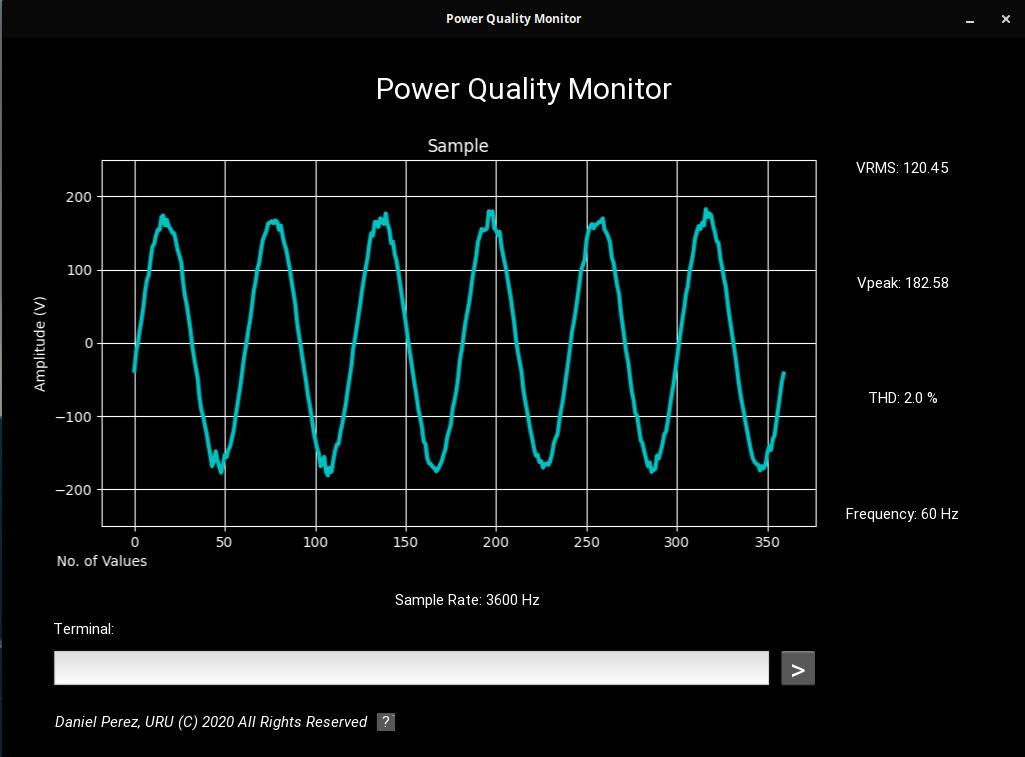}
\caption{6-cycle sample taken from Maracaibo, Venezuela}
\label{sample}
\end{figure}
\\As can be seen in figure \ref{sample}, there's distortion in the peaks of the sine wave. The distortion isn't present on the linear portions of the sine wave, therefore there's no clear indication that the peak distortion is due to noise. The distortion isn't really affecting the RMS value, because it is hovering around 120 VRMS, as it should according to Venezuela's national electrical code \cite{CEN}. Being that the main GUI has the ability of toggling between instantaneous and RMS $\frac{1}{2}$ data, we can see the corresponding RMS $\frac{1}{2}$ of these 6 cycles:
\begin{figure}[!ht]
\centering
\includegraphics[width=80mm]{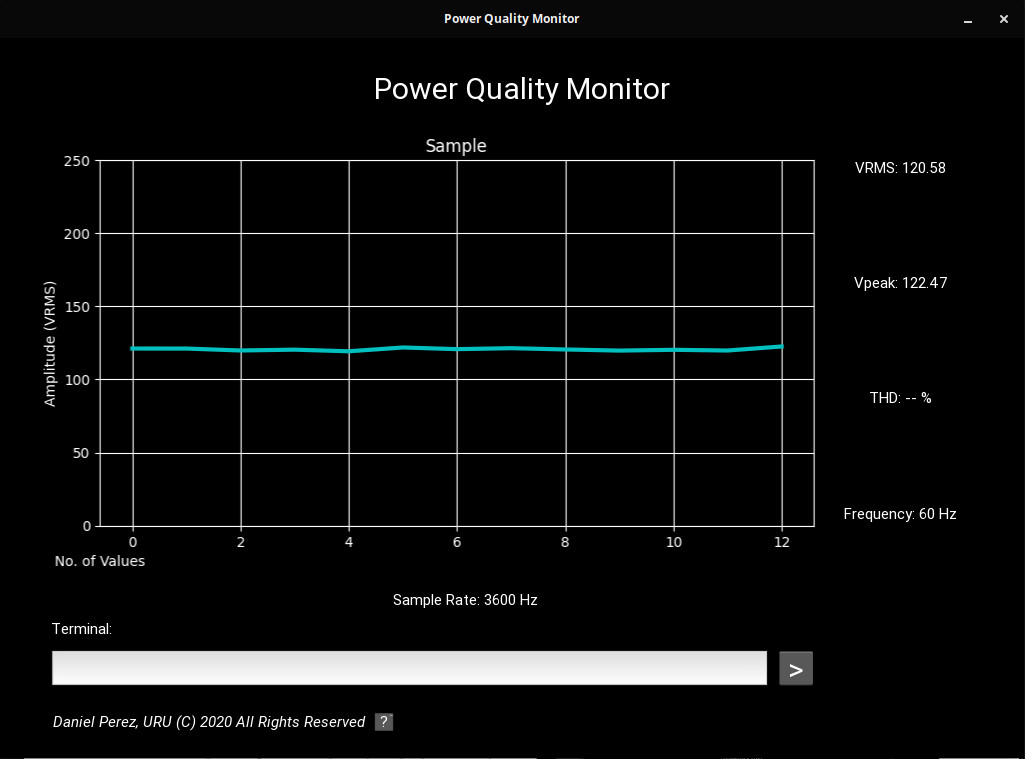}
\caption{RMS $\frac{1}{2}$ values of the 6-cycle sample}
\label{samplerms}
\end{figure}
\\The overall THD of the sample indicates that there's definitely presence of harmonics, but they aren't prominent and they might be conditioned by noise. In figure \ref{fftsample} the FFT of the sample can be seen:
\begin{figure}[!ht]
\centering
\includegraphics[width=85mm]{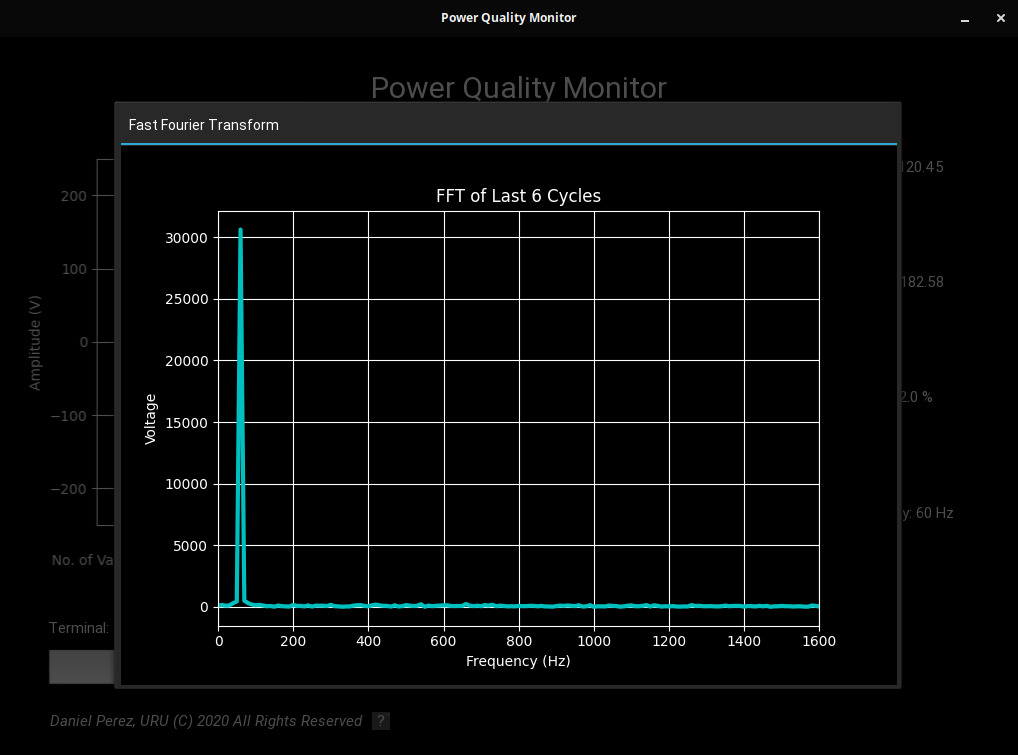}
\caption{Fast Fourier transform of the sample shown in figure \ref{sample}}
\label{fftsample}
\end{figure}
\\The vast majority of the energy is being transmitted at 60 Hz, but there's energy transmission occurring at practically all harmonics of the fundamental frequency. This scenario opens up the possibility that the readings are being affected by noise, being that there's no other prevalent spike in the spectrum. A comparison with a close to an ideal sine wave signal has to be drawn u-sing the same hardware, to determine whether or not noise is having a great impact. Also, a comparison with professional power quality equipment would be ideal to confirm whether the real THD value is close or not to what is being calculated by the system.\\
Lastly, a power quality analysis report can be seen in the following figure:
\begin{figure}[!ht]
\centering
\includegraphics[width=85mm]{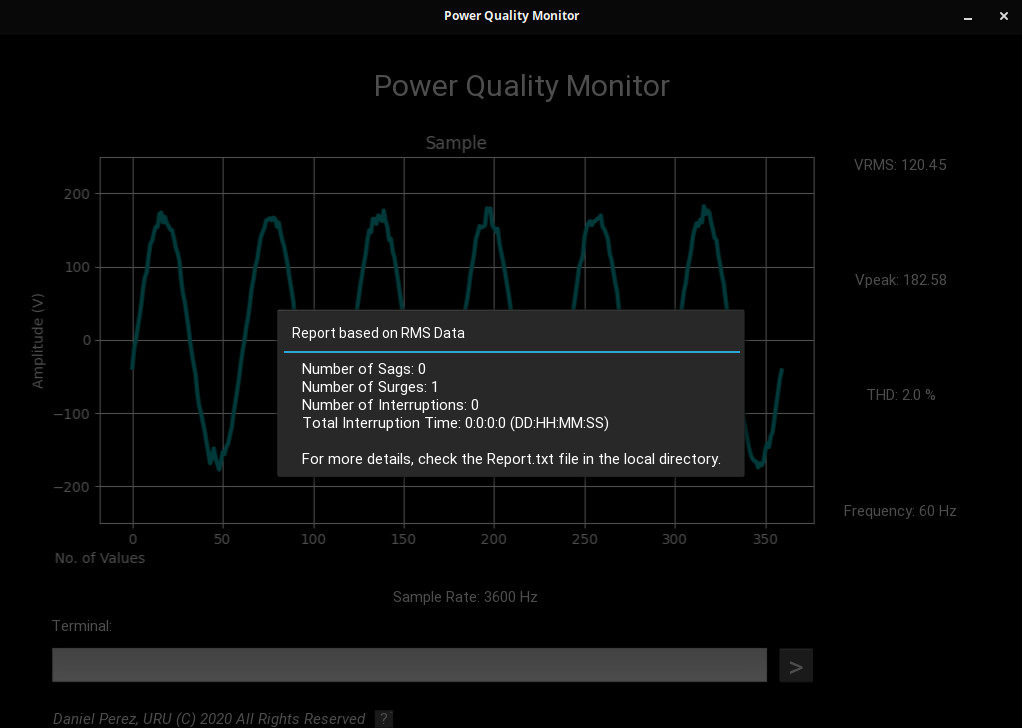}
\caption{Main stats of power quality analysis report of entire sample}
\label{report}
\end{figure}
\\The report only indicates that a voltage surge took place. By then inspecting the ``Report.txt'' file it was determined that the surge had a duration of only one semi-cycle. No voltage sags or interruptions occurred.

\section{Conclusions}
The system is capable of perceiving and storing the value changes that come with voltage sags and surges, so it can be used for their study. It can be also used for making a light power quality assessment. It requires inexpensive hardware and takes advantage of free software. It can be easily scaled to analyze Maracaibo's electrical system, having only as a setback the poor state of Internet connectivity in the city. Comparison with professional power quality equipment and with a sine wave from a signal generator is pending. It has great potential for expanding functionality.

\end{document}